\documentclass[twocolumn,floatfix,nofootinbib,10pt]{revtex4-1}
\usepackage{amssymb,amsmath}
\usepackage{graphicx,float}
\usepackage{indentfirst}
\newcommand{\be}{\begin{equation}}
\newcommand{\ee}{\end{equation}}

\newcommand{\ba}{\begin{eqnarray}}
\newcommand{\ea}{\end{eqnarray}}

\newcommand{\bx}{\vec{x}}
\newcommand{\e}[1]{\operatorname{e}^{#1}}
\usepackage{color}
\begin{document}
%
%
\title{Stability of the graviton Bose-Einstein condensate in the brane-world}
\author{R. Casadio}
\email{casadio@bo.infn.it}
\affiliation{Dipartimento di Fisica e Astronomia, Universit\`a di Bologna, 
via Irnerio 46, 40126 Bologna, Italy
\\
INFN, 
Sezione di Bologna, viale B.~Pichat 6, 40127 Bologna, Italy}
\author{Rold\~ao da Rocha}
\email{roldao.rocha@ufabc.edu.br} \affiliation{
CMCC,
Universidade Federal do ABC, 09210-580, Santo Andr\'e, SP, Brazil}


\begin{abstract}
{We consider a solution of the effective four-dimensional Einstein equations,
obtained from the general relativistic Schwarzschild metric through the principle of Minimal
Geometric Deformation (MGD).
Since the brane tension can, in general, introduce new singularities on a relativistic E\"otv\"os
brane model in the MGD framework, we require the absence of observed singularities, in order
to constrain the brane tension.
We then study the corresponding Bose-Einstein condensate (BEC) gravitational system and
determine the critical stability region of BEC MGD stellar configurations.
Finally, the critical stellar densities are shown to be related with critical points of the information
entropy. 
}
\end{abstract}
\maketitle
\flushbottom
%
%
%
\section{Introduction}
Several aspects of black hole physics have been recently studied, by considering
black holes as Bose-Einstein condensates (BEC) of a large number $N$ of weakly
interacting, long-wavelength, gravitons close to a critical
point~\cite{Dvali:2011aa,Dvali:2012gb,Dvali:2012rt}.
This paradigm has the merit to  directly interconnect black hole physics
to the study of critical phenomena, where quantum effects are
relevant at critical points, even for a macroscopic number $N$ of
particles~\cite{Flassig:2012re}. 
Although black holes are non-perturbative gravitational objects,
the effective quantum field theory of gravitons that describes them can
still be weakly coupled, due to large collective effects~\cite{Dvali:2012rt,hooft,witten}. 
Black hole features that cannot be recovered in a standard semiclassical approach of
gravity may then be encoded by the quantum state of the critical BEC~\cite{Casadio:2015lis,Casadio:2015jha},
with the semiclassical regime obtained as a particular limit for $N\to\infty$. 
Moreover, describing black holes by a condensate of long-wavelength gravitons
generates a self-sustained system, whose size equals the standard Schwarzschild radius
and the gravitons are maximally packed~\cite{Dvali:2011aa,Dvali:2012gb,Dvali:2012rt,Casadio:2015lis}.
A quantum field-theoretical analysis also clarified the relation between the emerging
geometry of spacetime and the quantum theory~\cite{Hofmann:2014jya}. 
\par
Brane-world models are effective five-dimensional (5D)
phenomenological realisations of the Ho$\check{\rm r}$ava-Witten domain wall
solutions~\cite{HW}, when moduli effects, engendered from the remaining extra dimensions, 
may be disregarded~\cite{Antoniadis:1998ig,maartens}. 
The brane self-gravity is identified by the brane tension $\sigma$, and the effective
four-dimensional (4D) geometry, due to a compact stellar distribution, can be achieved by a
Minimal Geometric Deformation (MGD) of the standard Schwarzschild solution in
General Relativity (GR)~\cite{ovalle2007,Ovalle:2010zc,Ovalle:2013vna,covalle1,covalle2}.
The MGD method ensures that this brane-world effective gravitational solution
reduces to the standard Schwarzschild solution, in the limit of infinite brane
tension $\sigma^{-1}\to 0$.
Therefore the MGD is a framework that provides corrections to GR,  
controlled by a parameter $\zeta$, that is a function of the stellar distribution effective radius and the brane tension.
\par 
Finally, we recall that a harmonic black hole model was recently introduced~\cite{qmm},
which can be viewed as an explicit realisation of a BEC of gravitons, with a regular interior.
The energy density in this model is obtained from a three-dimensional harmonic potential,
``cut''  around the horizon size in order to accommodate for the continuum spectrum of 
scattering modes, and the Hawking radiation. 
Afterwards, this model was ameliorated by instead considering the P\"oschl-Teller
potential~\cite{Muck:2014kea}, which naturally contains a continuum spectrum above
the bound states, contrary to the harmonic oscillator. 
\par
We shall here employ this last model to study a MGD BEC black hole and analyse
its critical stable density, from the point of view of the information entropy~\cite{glst,glsow},
and statistical mechanics~\cite{Bernardini:2016hvx}. 
The information entropy has been applied to a variety of settings, and the stability of
self-gravitating compact objects was already reported in Refs.~\cite{glst,Gleiser:2015rwa}. 
In particular, Newtonian polytropes, neutron stars, and boson stars 
were studied in Ref.~\cite{Gleiser:2015rwa}.
The information entropy is well-known to measure the underlying shape complexity of spatially localised configurations~\cite{glst,glsow}.
The less information involved in the modes that comprise a physical system, the smaller entropic information
is required to represent the same physical system.
The energy density
is the main ingredient to compute the information entropy.  
In this framework, the critical stable density of a BEC MGD black hole will be here
studied, by relating the stellar distribution conditional entropy and its central critical density.
In other words, the conditional entropy will be used to study the gravitational stability.
\par
This work is organised as follows:
we review the MGD procedure in Section~\ref{sMGD}, and the BEC description
of a MGD black hole is employed to establish a bound for the brane tension
of a E\"otv\"os brane-world model in Section~\ref{rel};
Section~\ref{IV} is devoted to establish the interplay between the critical
point in the stellar stability and the critical point of the conditional entropy
in a BEC MGD self-gravitating system scenario;
finally, we comment on our findings in Section~\ref{V}.
\section{minimal geometric deformation}
\label{sMGD}
The MGD approach is designed to produce brane-world
corrections to standard GR solutions, hence it is a suitable method to obtain
inhomogeneous, spherically symmetric, stellar distributions that are physically
admissible in the brane-world~\cite{covalle2,darkstars}.
For example, the bound $\sigma \gtrapprox  5\times10^6 \;{\rm MeV^4}$
for the brane tension was obtained from the MGD in Ref.~\cite{Casadio:2015jva}. 
The MGD was originally applied in order to deform the standard Schwarzschild
solution~\cite{ovalle2007,covalle1,covalle2} and describe the 4D geometry
of a brane stellar distribution.
Moreover, the MGD paved the way for interesting developments concerning 5D
black string solutions of 5D Einstein equations~\cite{Casadio:2013uma} in E\"otv\"os variable brane tension 
models~\cite{Bazeia:2013bqa,Bazeia:2014tua}. 
\par
The method relies on the effective Einstein equations on the brane~\cite{GCGR},
\begin{equation}
\label{gmunu}
R_{\mu\nu}-\frac12Rg_{\mu\nu}+\Lambda\, g_{\mu\nu}-\tilde{T}_{\mu\nu}=0
\ ,
\end{equation}
where the effective energy-momentum tensor is given by
\be
\tilde{T}_{\mu\nu}=T_{\mu\nu}+{\cal E}_{\mu\nu}+\frac{1}{\sigma}\,S_{\mu\nu}
\ ,
\label{Teff}
\ee
which contains the usual stress tensor $T_{\mu\nu}$ of brane matter (with
four-velocity $u^\mu$), and the (non-local) Weyl and high energy Kaluza-Klein
corrections $\cal{E}_{\mu\nu}$ and $S_{\mu\nu}$.
The Weyl tensor can be further decomposed as
\be
\label{emunu}
{\cal E}_{\mu\nu}
=
\frac{6}{\sigma}\left[{\cal U}\left(\frac{1}{3}\,h_{\mu\nu}+u_\mu\,u_\nu\right)
\!+\!{\cal P}_{\mu\nu}\!+\!{\cal Q}_{(\mu}\,u_{\nu)}\right]
\,,
\ee
where $h_{\mu\nu}=g_{\mu\nu}-u_\mu u_\nu$ denotes the induced spatial
metric, ${\cal P}_{\mu\nu}$ is the anisotropic stress, ${\cal U}$ stands
for the Weyl bulk scalar, and ${\cal Q}_\mu$ denotes the energy flux field.
\par
One then considers the general spherically symmetric metric, 
\begin{equation}
\label{abr}
ds^{2} = A(r) \,dt^{2} - \frac{dr^{2}}{B(r)} - r^{2} \,d\Omega^{2}
\ ,
\end{equation}
in the effective equations~\eqref{gmunu}. 
Any deformation of this static metric, with respect to a GR solution, must be caused
by 5D bulk effects, in a brane-world scenario.
Particularly, the radial component outside a compact stellar distribution, of average radius $r=R$, 
turns out to be given by~\cite{covalle1,covalle2} 
\begin{eqnarray}
\label{g11vaccum}
B_+(r)
=
{1-\frac{2\,M}{r}}
+\zeta\,e^{-I}, 
\end{eqnarray}
where  
\be
\!I(r)\!=\!
\int^r_{R}\!\!\!
\left({\!\frac{AA''}{A'^{2}}\!+\!\frac{A'^{2}}{A^2}\!-\!1\!+\!\frac{2A'}{rA}\!+\!\frac{1}{r^{2}}}\!\right)
\!\!\left(\frac{2}{r}\!+\!{\frac{A'}{2A}}\right)^{-1}
\!\!\!\!d\bar r
\ ,
\label{I}
\ee
where primes denote derivatives with respect to $r$.
The parameter $\zeta$ describes the deformation induced onto the vacuum by bulk effects,
evaluated at the surface of the stellar distribution.
Therefore, $\zeta$ contains all relevant information of a Weyl fluid on the brane~\cite{Casadio:2015jva}. 
The matching conditions with the inner star metric then determine the outer metric
for $r>R$~\cite{ovalle2007,Casadio:2013uma}.
In particular, if one considers the standard Schwarzschild metric, the deformed outer metric
components read~\cite{covalle1}
\begin{subequations}
\ba
\label{nu}
\!\!\!\!\!\!A_+(r)
&=&
1-\frac{2\,M}{r}
\ ,
\\
\!\!\!\!\!\!B_+(r)
&=&
\left(1-\frac{2\,M}{r}\right)
\left[1+{\zeta}\frac{\ell}{r}\left({1-\frac{3\,M}{2\,r}}\right)^{-1}\,\right]
\ ,
\label{mu}
\ea
\end{subequations} 
where $\ell$ is a length given by~\footnote{The deformation around the star surface is negative,
in order to prevent a negative pressure  for a solid crust~\cite{darkstars}.} 
\begin{equation}
\label{L}
\ell
\equiv
R
\left(1-\frac{2M}{R}\right)^{-1}
\left(1-\frac{3M}{2R}\right)
\ .
\end{equation}
{This metric has two event horizons where $B_+=0$:
one is the usual Schwarzschild horizon, $r_s = 2\,M$, and the second horizon is at
$r_2=\frac{3M}{2} - \zeta\,\ell$.
The expression of $\zeta$ was previously derived~\cite{ovalle2007,covalle1},
\begin{eqnarray}
\label{betasigma}
\zeta(\sigma,R)
\approx
-\frac{0.275}{R^2\,\sigma}
\ ,
\label{c0}
\end{eqnarray}
and the GR limit $\zeta\sim \sigma^{-1}\to0$ implies that  $r_2< r_s$.
One can therefore conclude that the gravitational field around the compact star
is weaker than in GR.
}
\section{BEC and MGD: a brane tension bound}
\label{rel}
In order to study BEC black holes with the MGD methods, let us start from
the Klein-Gordon equation for a scalar field $\Psi$~\cite{Muck:2014kea}
\begin{equation}
\label{rel:KG}
\left\{\left[i\,\hbar\,\partial_t -V(\bx)\right]^2
+\hbar^2 \,\nabla^2
-\left[\mu+S(\bx)\right]^2
\right\}
\Psi(t,\bx)
=
0
\ ,
\end{equation}
where  $\mu$ denotes the rest mass and one included the time-independent vector
and scalar potentials $V(\bx)$ and $S(\bx)$.
Writing $\Psi(t,\bx) = \e{-i\,{\varpi}\, t/\hbar} \Psi(\bx)$ and assuming
$S=V$ yield 
\begin{equation}
\label{schrodinger}
\left[ -\frac{\hbar^2}{2\,({\varpi}+\mu)}\, \nabla^2
+V
-\frac{1}{2}\,(\varpi - \mu) \right] \Psi(\bx) = 0
\ ,
\end{equation}
which is just a Schr\"odinger equation with $m= \varpi+\mu~,$ and
$E= \frac{1}{2} \left(\varpi-\mu\right)$.
It represents the relativistic dispersion relation $\varpi^2 = \hbar^2 k^2+\mu^2$,
and we shall in particular  consider the spherically symmetric
P\"oschl-Teller potential~\cite{Muck:2014kea}
\be
V
=
-\frac{3\,\mu}{\cosh(\mu\,r/\hbar)}
\ ,
\ee
for which one can find explicit solutions for $\Psi=\Psi(r)$ and compute the
corresponding energy density.
In fact, this graviton BEC can be macroscopically modelled by an anisotropic fluid,
with local energy-momentum tensor of the form 
\begin{equation}
\label{energym}
T^{\mu\nu}
=
\left(p_\parallel-p_\perp\right) v^\mu v^\nu+\left(\varepsilon+p_\perp\right)
u^\mu u^\nu+p_\perp\, g^{\mu\nu}
\ ,
\end{equation} 
where $u^\mu\,u_\mu=-1=-v^\mu\,v_\mu$, and $u^\mu\,v_\mu=0$,
$\varepsilon$ is the energy density, $p_\perp$ and $p_\parallel$ are the pressures
perpendicular and parallel to the space-like vector $v^\mu$. 
For the static, spherically symmetric metric in Eq. (\ref{abr}),
one has 
\begin{equation}
\label{fluid1}
	u^\mu = \left(-A^{-1/2},0,0,0\right)^\intercal,\quad 
	v^\mu = \left(0,B^{1/2} ,0,0\right)^\intercal
	\ .
\end{equation}
One can then introduce the (quasilocal) Misner-Sharp mass function,
\begin{equation}
\label{mass1}
M(r) = 4\pi \int_0^r  {\bar{r}}^2 \varepsilon(\bar{r})\;d \bar{r}
\ ,
\end{equation}
which represents the total energy within a stellar distribution of radius $r$,
and find the radial component of the metric~\cite{Muck:2014kea}
\be\label{brho}
B(\rho)
= 
1+\frac{1}{\rho} \tanh^3(\nu \rho)\left[3\tanh^2(\nu\rho)-5\right]
\ ,
\ee
where $\rho= {r}/{M} = {2\,r}/{r_s}$ and $\nu =M\,\mu/\hbar$, with $r_s$ the
gravitational radius of the total Misner-Sharp mass $M$ like in the previous Section.
\par
\begin{figure}[t]
\includegraphics[scale=0.6]{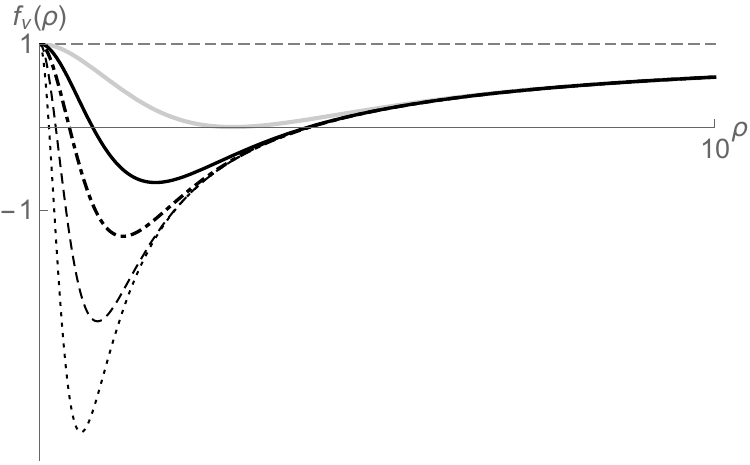}
\caption{Plot of $B_\nu(\rho)$ in Eq.~\eqref{Brho},
for $\nu = 0$ (gray dashed line);
$\nu = 0.3$ (thick gray line);
$\nu = 0.5$ (thick black line);
$\nu = \nu_\ast$ (black dot-dashed line);
$\nu = 1$ (black dashed line);
$\nu = 1.4$ (dotted line). 
}\label{fig1}
\end{figure}

The fluid description by means of Eq.~\eqref{energym} makes it now possible
to apply the MGD approach to the above metric.
In particular, the graviton BEC model of Ref.~\cite{Muck:2014kea} considers the
equation of state $\varepsilon + p_\parallel =0$.
Upon assuming for the temporal component of the metric the usual Schwarzschild
form~\eqref{nu}, the effective energy-momentum tensor $\tilde T_{\mu\nu}$
in Eq.~\eqref{Teff} yields the effective energy density and pressure~\cite{darkstars}
\begin{eqnarray}
\label{density1}
\!\!\!\!\!\!\tilde\varepsilon
&=&
-\frac{(9-14cr^2+c^2r^4)}{7\pi(1+cr^2)^4}\,
\delta+\frac{c\,(c^2r^4\!+\!2cr^2\!+9)}{7\pi(1+cr^2)^3},
\\
\!\!\!\!\!\!\tilde p
&=&
-\frac{4(2c^2r^4-9cr^2+1)}{7\pi(1+cr^2)^4}\,
\delta+\frac{2c(c^2r^4+7cr^2+2)}{7\pi(1+cr^2)^3},
\quad
\label{pressure1}
\end{eqnarray}
where $c\simeq 0.275/R^2$.
In the next Section, Eqs.~\eqref{density1} and \eqref{pressure1}
will also be used to compute the effective energy density as the temporal component of
the effective energy-momentum tensor~\eqref{Teff}, namely
\begin{equation}
\label{effd}
\tilde\epsilon(r)=\tilde T^{00}(r)
\ .
\end{equation}
In the above expressions, the brane-world corrections are given by the terms proportional to 
\be
\label{deltaf}
\delta(\sigma)
=
\frac{f^{*}_R}{\sigma\,R^2}\frac{7(1+cR^2)^2(1+9cR^2)}{16\,R^2(7+2cR^2)}
+
\mathcal{O}(\sigma^{-2})
\ ,
\ee
with
\begin{eqnarray}
f^{\ast}_R
&=&
\frac{4}{49\,\pi }
\left[\frac{80\,\mathrm{arctan}({y}^{1/2})}{(1+y)^{2}(3y+1){y^{1/2}}}
\right.
\nonumber
\\
&&
\left.
\qquad
+ \frac{3y^{4}+41y^{3}+25y^{2}-589y-240}{3(1+y)^{4}(1+3y)}
\right]
\ ,
\label{mgd}
\end{eqnarray}
for $y=c\,R^2$~\cite{darkstars}.
Finally, in the limiting case $R=r_s$, which we assume describes the BEC
black hole, the deformed radial metric component to leading order in
$\sigma^{-1}$ is given by
\begin{equation}
\frac{B_\nu(\rho)}{B(\rho)}
=
1-
\frac{2\,c_0}{\sigma\!\left[\rho-\frac{3}{4} \tanh^3(\nu \rho)\right]
\!\!\left[ 5-3\tanh^2(\nu\rho)\right]}
\ ,
\label{Brho}
\end{equation}
where $c_0\simeq 0.275$.
Fig.~\ref{fig1} shows plots of $B_\nu(\rho)$ for various values of $\nu$.
It is clear that, for increasing values of $\nu$, this black hole model
rapidly approaches the Schwarzschild black hole.
This figure can be compared to Fig.~1 in Ref.~\cite{Muck:2014kea}
for similar parameters.

\par
For any $\nu$, the metric component $B_\nu(\rho)$ has a single local minimum at
$\rho_\ast = a_\ast/\nu$, where $a_\ast \approx 1.031.$
Writing $B_\nu(\rho_\ast) = 1-{\nu}/{\nu_\ast}$, 
with $\nu_\ast \approx 0.694$,
the condition for the existence of an event horizon is $\nu > \nu_\ast$. 
The case $\nu=\nu_\ast$ is extremal \cite{Muck:2014kea}.
\subsection{Variable tension model}
{A more realistic model can be implemented by considering an E\"otv\"os
variable tension brane~\cite{gly2,Abdalla:2009pg}.
Essentially, the E\"otv\"os law states that the (fluid) membrane tension
depends on the temperature as
\be
\sigma
=
\xi
\left(T_{\rm crit}-T\right)
\ ,
\label{TERM0}
\ee
where $\xi$ is a positive constant and $T_{\rm crit}$ a critical temperature that 
determines the ceasing of the membrane existence.
The tension variation is now expressed in terms of the (cosmic) time,
instead of the temperature.
Indeed, the Universe cools down as it expands.
The cosmic microwave background
indicates $T\sim a^{-1}$, where $a=a(t)$ denotes the FLRW Universe
expansion factor~\cite{gly2,Bazeia:2013bqa,Bazeia:2014tua}, 
in agreement with the standard cosmological model.
Eq.~(\ref{TERM0}) then yields
\be
\sigma(t)=\sigma_0\left(1-\frac{a_{\rm min}}{a(t)}\right)
\ ,
\label{opa11}
\ee
where $\sigma_0$ is a constant related to the 4D coupling constants~\cite{gly2},
and $a_{\rm min}$ is the minimum scale factor, below which  the brane does cease to exist.
\par
{Along each definite phase, in the Universe evolution, brane
tension changes are just perceptible across cosmic time scales.}
{Regarding a de Sitter (dS) brane, the variable brane tension reads
 $\sigma(t)\sim1-e^{-\chi t},$ for $\chi>0$ \cite{RE41}, being admissible from a phenomenological point of view~\cite{maartens,GCGR}, 
predicting a variable Newton coupling constant $G\sim \sigma(t)$. 
{In Fig.~\ref{F2} and~\ref{F3}, we display a refinement of Fig.~\ref{fig1}, along 
the time scale, 
which takes into account an E\"otv\"os brane variable tension,
according to the law~\eqref{opa11}.} 
\begin{figure}[t]
\begin{center}
\includegraphics[scale=0.6]{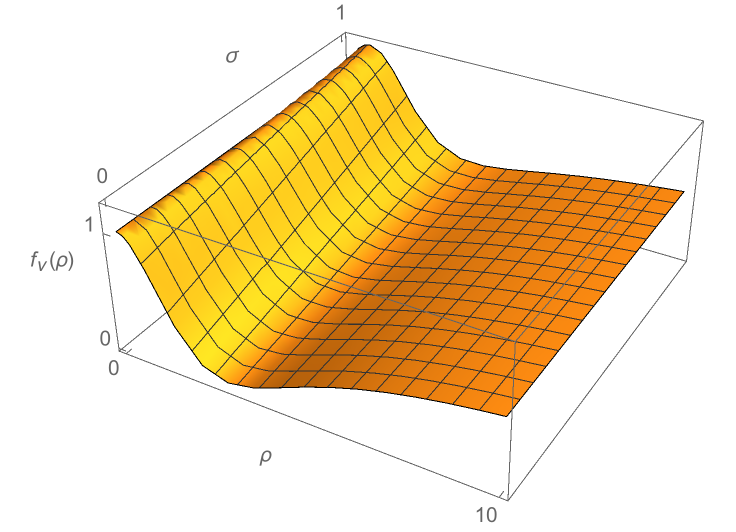}
\caption{Plot of $B_\nu(\rho)$ for $\nu = 0.3$, in the era dominated by non-relativistic
matter, as a function of the E\"otv\"os brane tension $\sigma(t)$
and the radial coordinate $\rho={r}/{M}$. The cosmological time is normalised according to Eq.~(\ref{opa11}).}
\label{F2}
\end{center}
\end{figure}
\begin{figure}[t]
\begin{center}
\includegraphics[scale=0.6]{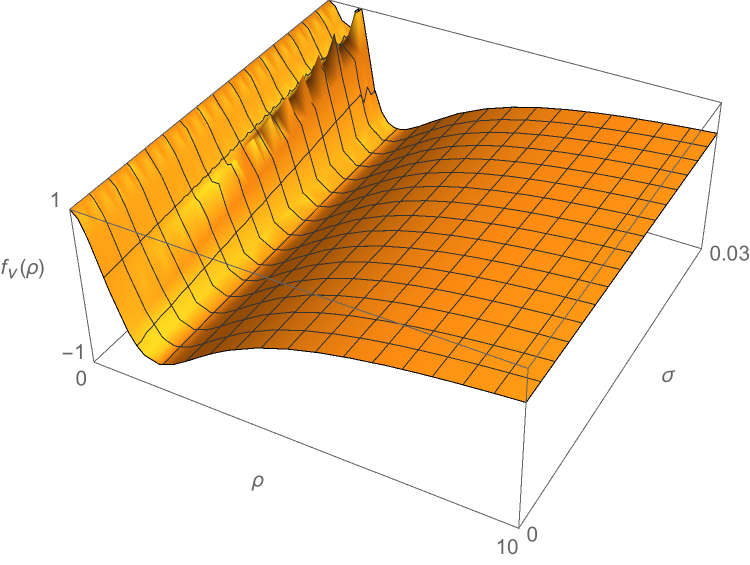}
\caption{Plot of $B_\nu(\rho)$ for $\nu = 0.5$, in the era dominated by the cosmological constant,
as a function of the E\"otv\"os brane tension $\sigma(t)$ and the radial coordinate $\rho={r}/{M}$.
The cosmological time is normalised according to Eq.~(\ref{opa11}).}
 \label{F3}
\end{center}
\end{figure}
\par
{Since we do not observe extra singularities up to the standard Schwazschild-like one,
the brane tension is constrained by the bound:}
\begin{equation}
\sigma\gtrsim 3.18 \times 10^6 {\rm MeV}^4
\ .
\end{equation}
{This limit is much stronger than the cosmological nucleosynthesis constraint,
and also better than the one provided by the limits of the MGD parameter
$\zeta$~\cite{Casadio:2015jva}.}
\section{Stability analysis of the MGD BEC}
\label{IV}
Having computed a bound on the brane tension for the MGD BEC,
we can now employ the conditional entropy in order to obtain stability bounds 
of critical points of the MGD BEC stellar distribution density.  
The entropic information, realised by the  conditional entropy, was utilised 
to scrutinise the stability of physical systems~\cite{glsow}, and proved to be based on statistical
mechanics grounds in Ref.~\cite{Bernardini:2016hvx}. 
Indeed, physical systems have classical field configurations defined as a
critical points of the classical action.
In a semiclassical approximation, these configurations can then be thought as critical
points of the effective action. Critical points of conditional entropy correlated
with the most stable configurations, in the context of information
entropy~\cite{Correa:2015lla,Correa:2016pgr}. 
Physical systems states, with larger information entropy, either need a larger
amount of energy to be produced, or are more scarcely observed -- or detected --
than their configurationally stable analog states,
or even both~\cite{Bernardini:2016hvx,Bernardini:2016qit}. 
The conditional entropy is based upon the information entropy and has
critical points of stability that can comprise configurations that provide
the best compression of informational nature in the system. 
\par
In order to compute the conditional entropy for the MGD BEC stellar
distribution, let us start by calculating the spatial Fourier transform of the
energy density $\tilde\epsilon=\tilde\epsilon(r)$ in Eq.~\eqref{effd},
\begin{equation}  
\epsilon(\omega)
=
(2\pi)^{-1/2}\lim_{n\to\infty}\int_{-n}^n
\tilde\epsilon(r)\,e^{i\omega r}\, dr
\ ,
\label{collectivecoordinates}
\end{equation}
where $r$ is again the radial coordinate.
Employing the effective energy density appears quite natural, since this
quantity effectively describes the spatially localised BEC in the brane-world,
also including the physics and boundary conditions that determine the stellar distribution.
These Fourier components mimic the collective coordinates~\cite{Bernardini:2016hvx}.
The structure factor, defined by
$s_n = \frac1n\sum_{k=1}^n\;\langle\;\epsilon(\omega_k)^*\,\epsilon(\omega_k)\;\rangle$,
normalises the correlation of collective coordinates, and defines the discrete modal fraction 
\begin{equation}
f(\omega_n)=\frac{1}{n\,s_n}{\langle\; \epsilon^*(\omega_n)\,\epsilon(\omega_n) \;\rangle}
\ .
\label{collective1}
\end{equation}
The structure factor probes fluctuations in the energy density~\cite{Bernardini:2016hvx},
as the energy density operator fluctuates among system configurations.
In the limit $n \to \infty$, the discrete modal fraction~\cite{glst} can be expressed as the ratio
between the correlation of collective coordinates and the structure factor~\cite{Bernardini:2016hvx},
 \begin{equation}
 f(\omega)
 \equiv
 \lim_{n\to\infty}f(\omega_n)
 =
 \frac{\langle\;\left\vert \epsilon(\omega)\right\vert ^{2}\rangle}{{\displaystyle{\lim_{n\to\infty}\int_{-n}^{n}}}
 d\omega\;\langle\;\left\vert \epsilon(\omega)\right\vert ^{2}\rangle \,}
 \ .
 \label{collective}
 \end{equation}
Now, by denoting with $f_{\rm max}(\omega)$ the maximum modal fraction,
define $\mathring{f}(\omega)=f(\omega)/f_{\rm max}(\omega)$ \cite{glst}.
The conditional entropy computed in the lattice approach then reads~\cite{glst,glsow}  
\begin{equation}
S_c[f]
=
-\lim_{n\to\infty}
\int_{-n}^{n}d\omega
\, \sigma(\omega)\,\ ,
\label{conditional}
\end{equation} 
where $\sigma(\omega) = \mathring{f}(\omega)\ln\mathring{f}(\omega)$ denotes
the conditional entropy density~\cite{Gleiser:2015rwa}.
Note that, since $\mathring{f}(\omega)$ is not a periodic function, the limits of
integration in Eq.~\eqref{conditional} are $\omega = \omega_{\rm min}=\pi/R$
(the lower limit) and $\omega\to\infty$ (the upper limit), as in Ref.~\cite{Gleiser:2015rwa}.
\par
The stability of self-gravitating compact objects by means of the configurational entropy
was previously investigated in Refs.~\cite{glst,Gleiser:2015rwa}. 
Eq.~\eqref{conditional} further measures the configurational, or informational, stability
of the system modes.
Taking the continuum limit of Shannon entropy is subtle and can miss some features
that are demanded for any definition compatible with the expected properties of an entropy.
As a matter of fact, the greater the information entropy, the more information,
encrypted in the system  modes, is compelled to constitute the complexity of
a spatially localised system~\cite{glst,glsow,Bernardini:2016hvx}.
\par
To perform a stability analysis of the MGD BEC distribution, let us recall that most
studies of black hole physics are based on the background classical geometry.
Therefore, the semi-classical approximation which accounts for small fluctuations
about the classical background does not include the quantum effects on the
background.
In fact, the classical geometry should be an appropriate limit of an effective quantum theory,
in which states, with large graviton occupation number, can be thought of as the main
constituents.
When such states are the ground-state, the gravitational field can be effectively
regarded as a BEC.  
\par
\begin{figure}[t]
\begin{center}
\includegraphics[width=3.07in]{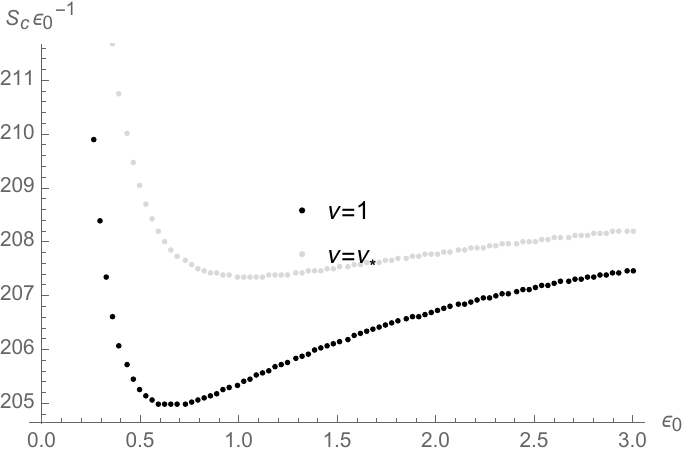}
\caption{Conditional entropy times the inverse of the BEC MGD black hole critical  central  density $\epsilon_0$, with respect to $\epsilon_0$.}
\label{f4}
\end{center}
\end{figure}
The equilibrium configurations can be derived in terms of the energy density
$\tilde\epsilon=\tilde\epsilon(r)$ in Eq.~\eqref{effd}.
The same density $\tilde\epsilon$ can then be employed, in order to compute
the conditional entropy $S_c$ as a function of the critical central density,
$\tilde\epsilon(0)\equiv \epsilon_0$, of the compact stellar distribution.
The conditional entropy~\eqref{conditional} is, in particular, obtained by means of the modal
fraction~\eqref{collective}, and the critical central density $\epsilon_0$ is found from
Eq.~\eqref{effd} to be $\epsilon_0\approx0.40925\, \delta(\sigma) + 0.11254/R^2$,
where $\delta(\sigma)$ is given by the expression in Eq.~\eqref{deltaf}.
The explicit expression of the conditional entropy turns out to be rather involved, so its
numerical estimate is displayed in Fig.~\ref{f4},
where $S_c$ is multiplied by the inverse central density to produce a quantity
that scales with dimensions of inverse mass. 
\par
The analysis of the conditional entropy for polytropes and neutron stars also
makes use of the fiducial value $\epsilon_c$ for the critical central density.  
In Ref.~\cite{Gleiser:2015rwa}, the critical points of the conditional entropy
were related to $\epsilon_c$ and to the Chandrasekhar mass likewise.
However, to our knowledge, there is no result in the literature that provides 
the analogue of the results in Fig.~\ref{f4}.
The two-fold reason for it is that the conditional entropy was successfully
analysed for boson stars formed by self-interacting scalar fields, and the stability
properties of boson stars were explored in their ground state~\cite{1998,Gleiser:2015rwa}.
Our results are therefore original and represent a potential reference
for further analysis of the stability of boson stars in the graviton
BEC scenario.
\section{Concluding Remarks}
\label{V}
Studying a MGD BEC in a E\"otv\"os brane-world scenario provides a strict bound
for the brane tension $\sigma\gtrsim 3.18 \times 10^6 {\rm MeV}^4,$ 
which is stronger than the bound determined by the study of cosmological nucleosynthesis. 
This bound is also more stringent than the one already obtained in the MGD
formalism in Ref.~\cite{Casadio:2015jva}.
Moreover, analysing the graviton BEC MGD black hole shows that it is important to take
into account quantum effects at the stability critical point, even for a macroscopic
number of particles.
The MGD BEC has an unlimited number of gapless modes, since it corresponds to
a  large occupation number $N$ at the stability critical point,
coinciding with a maximal packing.
The minimum value of the conditional entropy is seen, from Fig.~\ref{f4}, to vary according to 
the value of the parameter $\nu=M\mu/\hbar$, that was defined as a function of the
Klein-Gordon scalar field rest mass $\mu$ and the Misner-Sharp mass $M$.
In fact, for $\nu=\nu_*$, the critical conditional entropy occurs at the BEC MGD black hole
critical  central density $\epsilon_0\approx0.62$, whereas the value $\nu=1$ yields the minimal
conditional entropy to be at $\epsilon_0\approx1.03$. 
Based upon previous results in Ref.~\cite{Gleiser:2015rwa} for polytropes and neutron stars,
a bound on the critical BEC MGD black hole mass can still be obtained, by establishing values
of the star density, according to Fig.~\ref{f4}.
Nevertheless, the critical point is shifted in the star density, due to the MGD procedure
and the graviton BEC as well, to describe the stellar distribution.
\section*{Acknowledgments}
RC is partly supported by INFN grant FLAG. 
RdR thanks  FAPESP 2015/10270-0, CNPq 303293/2015-2 and CNPq 473326/2013-2 for partial financial support.


\begin{thebibliography}{99}

\bibitem{Dvali:2011aa}
G.~Dvali and C.~Gomez, 
{
  Fortsch. Phys.} {\bf 61} (2013) 742.

\bibitem{Dvali:2012gb}
G.~Dvali and C.~Gomez, 
   {Phys. Lett. B} {\bf
  716} (2012) 240.

\bibitem{Dvali:2012rt}
G.~Dvali and C.~Gomez, 
{Phys. Lett. B} {\bf 719} (2013) 419.

\bibitem{Flassig:2012re}
  D.~Flassig, A.~Pritzel and N.~Wintergerst,
  Phys.\ Rev.\ D {\bf 87} (2013)  084007.

\bibitem{hooft}  G. 't Hooft, 
Nucl. Phys. B {\bf 72}
(1974) 461.
\bibitem{witten} E. Witten, 
Nucl. Phys. B {\bf 160} (1979) 57.
%


\bibitem{Casadio:2015lis}
  R.~Casadio, A.~Giugno, O.~Micu and A.~Orlandi,
  Entropy {\bf 17} (2015) 6893.

\bibitem{Casadio:2015jha}
  R.~Casadio, R.~T.~Cavalcanti, A.~Giugno and J.~Mureika,
  Phys.\ Lett.\ B {\bf 760} (2016) 36

\bibitem{Hofmann:2014jya}
  S.~Hofmann and T.~Rug,
  Nucl.\ Phys.\ B {\bf 902} (2016) 302.
  
\bibitem{HW} P. Ho{\v r}ava and E. Witten, Nuclear Physics B \textbf{460} (1996) 506.

\bibitem{Antoniadis:1998ig}
  I.~Antoniadis, N.~Arkani-Hamed, S.~Dimopoulos and G.~R.~Dvali,
  Phys.\ Lett.\ B {\bf 436} (1998) 257.
  

\bibitem{maartens}
R. Maartens and K. Koyama, 
Living Rev. Rel. {\bf 13} (2010) 5.
%
%
\bibitem{ovalle2007}
J. Ovalle,
Int. J. Mod. Phys. D {\bf 18} (2009) 837.%


\bibitem{Ovalle:2010zc}
  J.~Ovalle,
  Mod.\ Phys.\ Lett.\ A {\bf 25} (2010) 3323.

\bibitem{Ovalle:2013vna}
  J.~Ovalle, F.~Linares, A.~Pasqua and A.~Sotomayor,
  Class.\ Quant.\ Grav.\  {\bf 30} (2013) 175019.

\bibitem{covalle1}
R.~Casadio and J.~Ovalle,
Phys.\ Lett.\ B {\bf 715} (2012) 251.
%
\bibitem{covalle2}
R.~Casadio and J.~Ovalle,
Gen. Relat. Grav. {\bf 46} (2014) 1669. 
%
%
\bibitem{qmm} R. Casadio and A. Orlandi, 
JHEP {\bf 1308} (2013) 025.

\bibitem{Muck:2014kea}
  W.~M\"uck and G.~Pozzo,
  JHEP {\bf 1405} (2014) 128.
%
%
%
\bibitem{glst} { M. Gleiser and N.
Stamatopoulos, Phys. Lett. B \textbf{713} (2012) 304. }


\bibitem{glsow} { M. Gleiser and D. Sowinski,
Phys. Lett. B \textbf{727} (2013) 272. }



\bibitem{Bernardini:2016hvx}
  A.~E.~Bernardini and R.~da Rocha,
  Phys. Lett. B {\bf 762} (2016) 107.

\bibitem{Gleiser:2015rwa} 
 M.~Gleiser and N.~Jiang,
 Phys.\ Rev.\ D {\bf 92} (2015) 044046.


\bibitem{darkstars}
J.~Ovalle, L. A.~Gergely and R.~Casadio
Class. Quant. Grav. 32 (2015) 045015.


\bibitem{Abdalla:2009pg}
  M.~C.~B.~Abdalla, J.~M.~Hoff da Silva and R.~da Rocha,
  Phys.\ Rev.\ D {\bf 80} (2009) 046003. 

\bibitem{Casadio:2013uma}
  R.~Casadio, J.~Ovalle and R.~da Rocha,
  Class.\ Quant.\ Grav.\  {\bf 31} (2014) 045016.

\bibitem{Bazeia:2013bqa}
  D.~Bazeia, J.~M.~Hoff da Silva and R.~da Rocha,
  Phys.\ Lett.\ B {\bf 721} (2013) 306.

\bibitem{Bazeia:2014tua}
  D.~Bazeia, J.~M.~Hoff da Silva and R.~da Rocha,
  Phys.\ Rev.\ D {\bf 90} (2014)   047902.

\bibitem{GCGR}
T. Shiromizu, K. Maeda and  M. Sasaki,
Phys. Rev. D {\bf 62} (2000) 043523.


\bibitem{Casadio:2015jva}
  R.~Casadio, J.~Ovalle and R.~da Rocha,
  Europhys.\ Lett.\  {\bf 110} (2015) 40003.





\bibitem{WILL} C. M. Will, {Living Rev. Relativity} {\bf 9} (2005) 3.


\bibitem{gly2} L. A. Gergely, {Phys. Rev. D} {\bf  79} (2009) 086007.

\bibitem{RE41} A. Campos, C. F. Sopuerta, 
Phys. Rev. D {\bf 63} (2001) 104012.


\bibitem{Correa:2015lla}
  R.~A.~C.~Correa, R.~da Rocha and A.~de Souza Dutra,
  Annals Phys.\  {\bf 359} (2015) 198.
   
\bibitem{Correa:2016pgr}
  R.~A.~C.~Correa, D.~M.~Dantas, C.~A.~S.~Almeida and R.~da Rocha,
  Phys.\ Lett.\ B {\bf 755} (2016) 358.


\bibitem{Bernardini:2016qit}
  A.~E.~Bernardini, N.~R.~F.~Braga and R.~da Rocha,
  arXiv:1609.01258 [hep-th].


\bibitem{1998} M. Gleiser, Phys. Rev. D {\bf 38} (1988) 2376 [E {\bf 39} (1989) 1257].


\end{thebibliography}
\end{document}